\documentstyle[12pt,epsf]{article}

\newcommand{\ds}{\displaystyle}
\newcommand{\ra}{\rightarrow}
\newcommand{\be}{\begin{equation}}
\newcommand{\ee}{\end{equation}}
\newcommand{\bea}{\begin{eqnarray}}
\newcommand{\eea}{\end{eqnarray}}
\newcommand{\ci}{\cite}
\newcommand{\bi}{\bibitem}
\newcommand{\nono}{\nonumber \\}

\newcommand{\dd}{\partial}

\newcommand{\s}{\sigma}

\newcommand{\rr}{\vec{\bf{r}}}
\newcommand{\kk}{\vec{\bf{k}}}
\newcommand{\qq}{\vec{\bf{q}}}
\newcommand{\cx}{\mbox{\bf{\^r}}}
\newcommand{\cq}{\mbox{\bf{\^q}}}
\newcommand{\ck}{\mbox{\bf{\^k}}}

\def\dal{\,\lower0.3ex\vbox{\hrule\hbox{\vrule\kern2pt\vbox{\kern4pt\kern4pt}
\kern2pt\vrule}\hrule}\,}

\def\s{\sigma}

\begin{document}

\title{\sl Diffraction of wave packets in space and time}
\vspace{1 true cm}
\author{G. K\"albermann$^*$
\\Soil and Water dept., Faculty of
Agriculture, Rehovot 76100, Israel}
\maketitle

\begin{abstract}

The phenomenon of wave packet diffraction in space and time 
is described.
It consists in a diffraction pattern whose spatial location
progresses with time.
The pattern is produced by wave packet quantum scattering off an
attractive or repulsive time independent potential.
An analytical formula for the pattern at $t\rightarrow\infty$
is derived both in one dimension and in three dimensions. 
The condition for the pattern to exist is developed.
The phenomenon is shown numerically and analytically 
for the Dirac equation in one dimension also.
An experiment for the verification of the phenomenon is described and simulated 
numerically.
\end{abstract}


$^*${\sl e-mail address: hope@vms.huji.ac.il}

\newpage

\section{\sl Wave packet diffraction in space and time}

The diffraction process is a cornerstone of wave scattering. 
Quantum mechanical scattering shows
diffractive phenomena in space, like nuclear diffractive scattering, 
and separately in time\ci{mosh}.
The combined effect of time dependent opening of slits
for plane monochromatic waves produces diffraction patterns in space and
time.\ci{zeilinger}
Both, diffraction in time, and diffraction in space and time
are of the utmost importance in testing the validity of
time dependent predictions of quantum mechanics.
Recent measurements of atomic wave diffraction\ci{prl},
 have indeed demonstrated that the {\sl diffraction in time},
process is supported by experiments.

The present work will show that there exists
a broad class of phenomena occurring in nonrelativistic and
relativistic quantum wave packet potential scattering
that behave as time dependent {\sl persistent} diffraction patterns.
The patterns are produced by a time independent potential.
In the present case, the pattern will be shown not to decay exponentially, like
for the {\sl diffraction in time} phenomenon, and may provide a
 tool to deepen our understanding of time dependent quantum processes, 
pertaining to the description of atomic (or other) beams by means of wave
packets.

The peak structure exists for all packets, but, it survives 
only for packets that are initially narrower than a value related to the
potential width. For wider packets, the peak structure
merges into a single peak.

The effect is named  {\sl wave packet diffraction in space and time}.
It will be shown that the structure exists for
repulsive and attractive potentials.
It also appears when the relativistically
invariant Dirac equation is used instead of the Schr\"odinger
equation.

The present work emerges from numerical investigations of 
one and two-dimensional wave packet scattering off an attractive potential 
that 
showed curious multiple peak structures resembling a diffraction pattern.
\ci{pra,jpa}
It was seen there that, wave packets that are 
narrower than the well width initially, backscatter as a wave train that is 
coherent and multiple peaked, a diffraction pattern that travels
backwards in one dimension and at large angles in two-dimensions.

The effect is absent for packets whose {\sl initial}
width was much larger than the potential extension. For this case, 
a smooth wave hump proceeds both forwards and backwards.
In section 2 we will treat the one dimensional case
analytically. The Dirac scattering in one dimension will be dealt
with in section 3. Section 4 generalizes the results of 
section 2 to three dimensions. In section 5 an experiment 
aimed at verifying the theoretical predictions will be
suggested and simulated. Section 6 summarizes the paper.

\section{\sl Wave packet diffraction in space and time in one dimension}

In order to develop an analytical formula for the
difrraction pattern, we resort
to the simplest possible case that can be dealt with almost completely
analytically. This is the example of a wave packet scattering off
a square well.

Consider a gaussian wave packet impinging from the left,
on a well located around the origin,

\bea\label{sqwell}
V(x)= -V_0~\Theta(w-|x|)
\eea

where, {2\sl w} is the width, $\Theta$ is the Heaviside function, 
and $V_0$, the depth.
The use of a square well facilitates the calculation, but, the results
do not depend on the sharpness of the well as evidenced by the numerical
results of \ci{pra}. (Theoretical arguments that support this
statement may be found below after eq.(\ref{asymp}).)
Using the results of \ci{bohm}, we can write immediately
both the reflected and transmitted packets. From this point
 on we take $\hbar= c = 1$.

\bea\label{psifull}
\psi(x,t)=~\int_{-\infty}^{\infty}\phi(k,x,t)~a(k,q_0)~
e^{-i~\frac{k^2}{2~m}~t}dk
\eea
where $\phi(k,x,t)$ is the stationary solution to the square well scattering
problem for each $\sl k$ and $\sl a(k,q_0)$ is the Fourier transform
amplitude for the initial wave function with average momentum $\sl q_0$.

Explicitly, in the backward direction that interests us here, for $\sl x<-w$

\bea\label{psi}
\phi(k,x,t)&=&D(k,k')~e^{i~k~(x-x_0)}+~F(k,k')~e^{-i~k~(x+x_0+2~w)}\nono
D(k,k')&=&1\nono
F(k,k')&=&\frac{E(k,k')}{A(k,k')}\nono
E(k,k')&=&-2~i~(k^2-k'^2)~sin(2~k'w)\nono
A(k,k')&=&(k+k')^2~e^{-2~i~k'~w}-(k-k')^2~e^{2~i~k'~w}
\eea

where 
\bea\label{form}
a(k,q_0)=e^{-\s^2~(k-q_0)^2}
\eea
with $\ds k'=\sqrt{k^2+2~m~|V_0|}$, and $\s$, the width parameter
of the packet.

The integral of the {\sl D} term can be performed explicitly.
We will call this contribution to the wave $\psi_{in}$.
For $t\ra\infty$ we have

\bea\label{in}
\psi_{in}&\approx&\sqrt{\frac{\pi}{\s^2+i~t/(2~m)}}~e^u\nono
u&=&i~\frac{m}{2~t}~(x-x_0)^2-\frac{m^2~\s^2}{t^2}~(x-x_0-q_0~t/m)^2
\eea

For long times, the oscillations in the exponent of the
term accompanying $F$ in the wave function are
extremely fast. The most important contributions come
from phases that are an extremum with respect to {\sl k}.
We will call the contribution to the
wave due to $F$, $\psi_{refl}$.
Using the saddle point method we find that, the extremal phase 
for $\psi_{refl}$ demands the real part of the momenta to be positive
for negative {\sl x}, namely $k_{extremal}\approx\frac{2\s^2~q_0-i~x}
{2~\s^2+i~t/(2~m)}$

We concentrate on the case $k'>>k$, the low energy regime, for
which the polychotomous effect was most visible.\ci{pra,jpa}

At $k\approx 0$ we have $F\approx-1$.
To order {\sl $\frac{1}{t}$}, for long times and distances $x>>x_0>>w$ 
using the properties of Gaussian integrals it is found that

\bea\label{refl}
\psi_{refl}&\approx&\sqrt{\frac{\pi}{\s^2+i~t/(2~m)}}~e^{\tilde u}\nono
{\tilde u}&=&i~\frac{m}{2~t}~(x+x_0)^2-\frac{m^2~\s^2}{t^2}~(x+x_0+q_0~t/m)^2
\eea

The full wave in the backward direction is
 obtained from the sum of the interfering waves of eqs.(\ref{in},\ref{refl}).
The amplitude of the wave for very long times becomes

\bea\label{asymp}
|\psi|&=&2~\sqrt{\frac{2~m~\pi}{t}}~e^{-z}~|sin\bigg(\frac{m~x}{t}(x_0
+2~i~\s^2~q_0)\bigg)|\nono
&=&2~\sqrt{\frac{2~m~\pi}{t}}~e^{-z}~\sqrt{sin^2(\frac{m~x~x_0}{t})+
sinh^2(\frac{2~\s^2~q_0~m~x}{t})}\nono
z&=&\s^2~\bigg(\frac{m^2~(x^2+x_0^2)}{t^2}+q_0^2\bigg)
\eea

This expression represents a diffraction pattern that travels in time
and persists.
This result hinges upon the value of {\sl F} as $k\ra0$.
However, this value is essentially the reflection amplitude for zero
momenta, that is equal to -1 regardless of the potential.
Therefore, the expression for $\psi$ above does not
depend on the choice of a square well.

Figure 1 shows a comparison of the expression above (without any
change in scale) to the numerical calculation of the reflected wave
for $t=1.2~10^7$, a huge time compared to the transit
time of the packet through the well.
The initial momentum is $q_0=0.4$, the well width $w=1$, the initial
position $x_0=-60$, and the mass parameter $m=40$. We chose a large mass
in order to be on the
nonrelativistic domain $\ds \frac{q_0}{m}=v=0.01$, for which the
Schr\"odinger equation is valid.
\begin{figure}
\epsffile{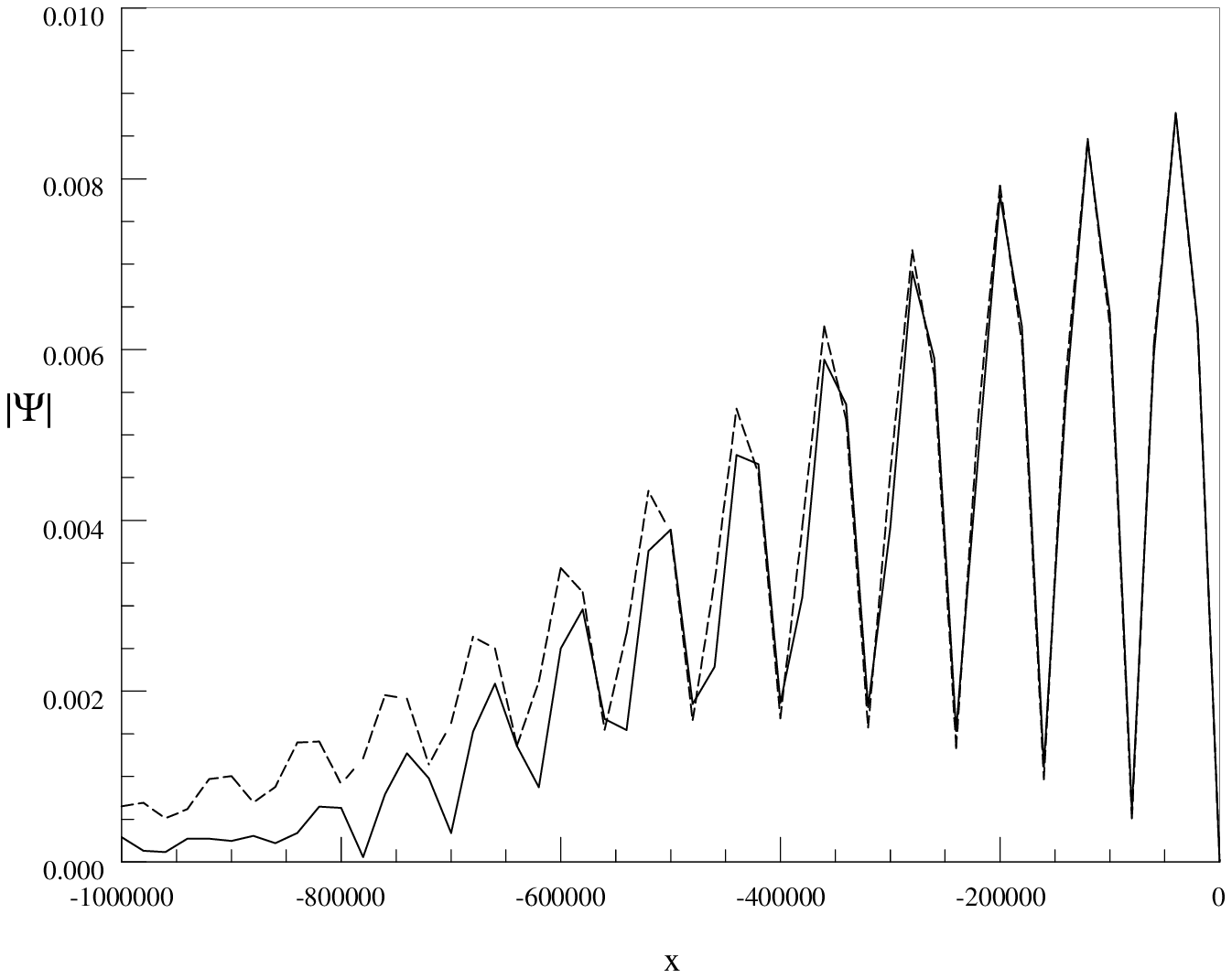}
\vsize=5 cm
\caption{\sl $|\Psi|$ as a function distance x, numerical calculation (solid
line) and theoretical expression of eq.(\ref{asymp}), (dashed line)
, at t=$1.2~10^7$.
The wave packet has average momentum q=0.4, mass m=40, 
width $\s=0.5$ and initial location $x_0$=-60. The well parameters are: width
w=1 and depth $V_0$=1.}
\label{fig1}
\end{figure}
The figure shows the theoretical expression and the numerical
calculation agree, despite the approximate treatment of
the $F$ term. For the longest distances, the approximation
is not valid. At these distances the approximation of $k\approx 0$
breaks down.
Due to the same time dependence of both the {\sl sin} and {\sl sinh}
pieces, it is quite clear that the pattern persists to infinite time.
The larger $x_0$, the cleaner the pattern, and more peaks appear
in the wave train. Clearly for $x_0\ra\infty$, the pattern disappears.
The smaller $q_0$, the more visible -less background-,
 the diffraction pattern is.

In order to understand the absence of diffractive structures for a 
broad packet we consider the imaginary term inside the $sin$ function in
eq.(\ref{asymp}).
We have to estimate the extension of the diffraction pattern that changes
with time. The momenta involved are $\ds k_{max}=\frac{m~|x_{max}|}{t}$
The constructive interference occurs whenever {\sl F} of eq.(\ref{psi})
 does not differ substantially from the value at $k=0$. At higher 
momenta {\sl F} is complex, the real part
diminishes and the imaginary part starts increasing. This
changes reduce the constructive part of the pattern, incoherent
pieces arise that blurr the multiple peak behavior.
The faster the drop of the real part of $-F$, the smaller
the range of momenta for which the formula above holds.
\begin{figure}
\epsffile{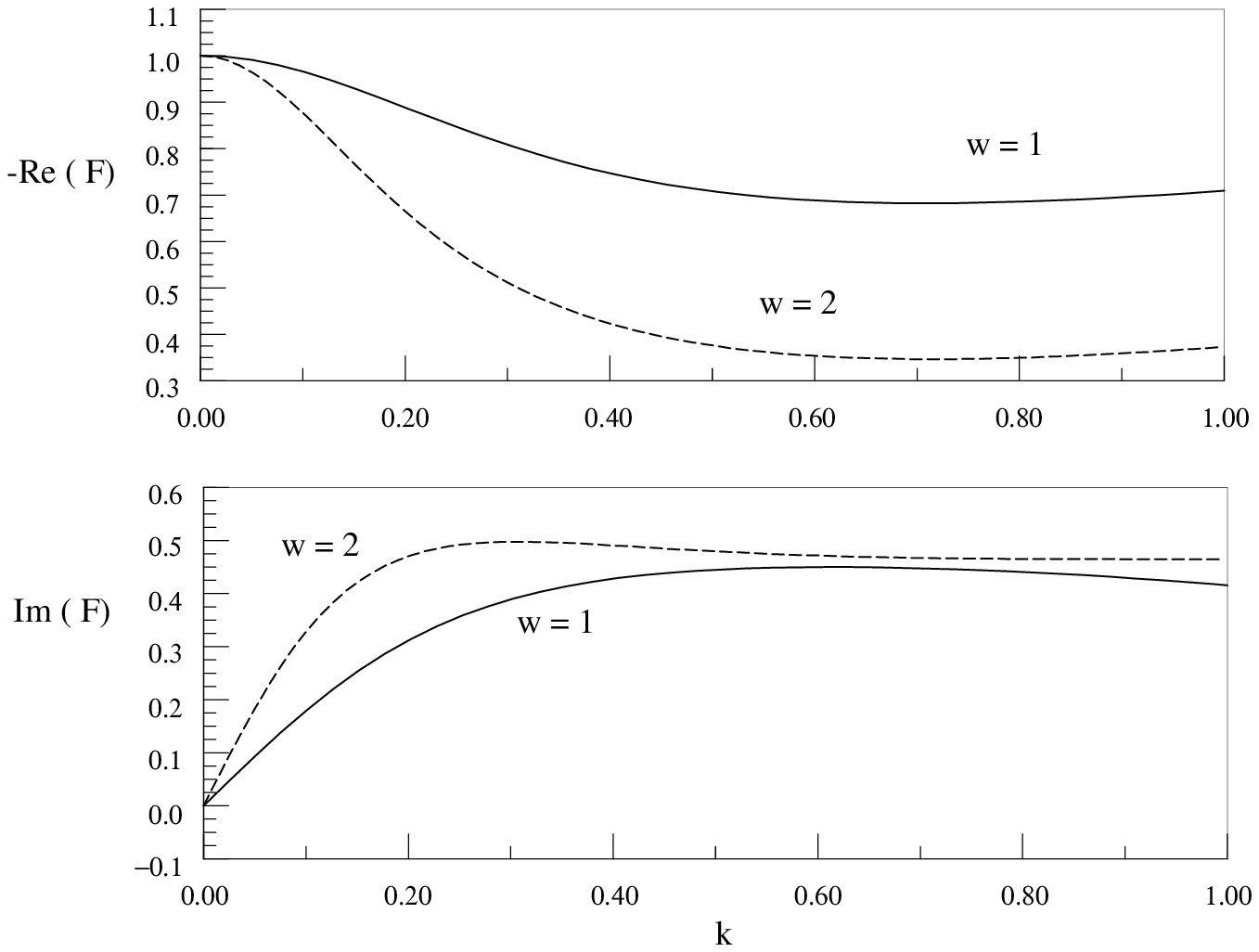}
\vsize=5 cm
\caption{\sl Negative Real and imaginary parts of the coefficient {\sl F}
of eq.(\ref{psi}) as a function of {\sl k} for the case
$\frac{k'_0~w}{2~\pi}=1.00658424209$, w=1, and twice this value, w=2}
\label{fig2}
\end{figure}
\begin{figure}
\epsffile{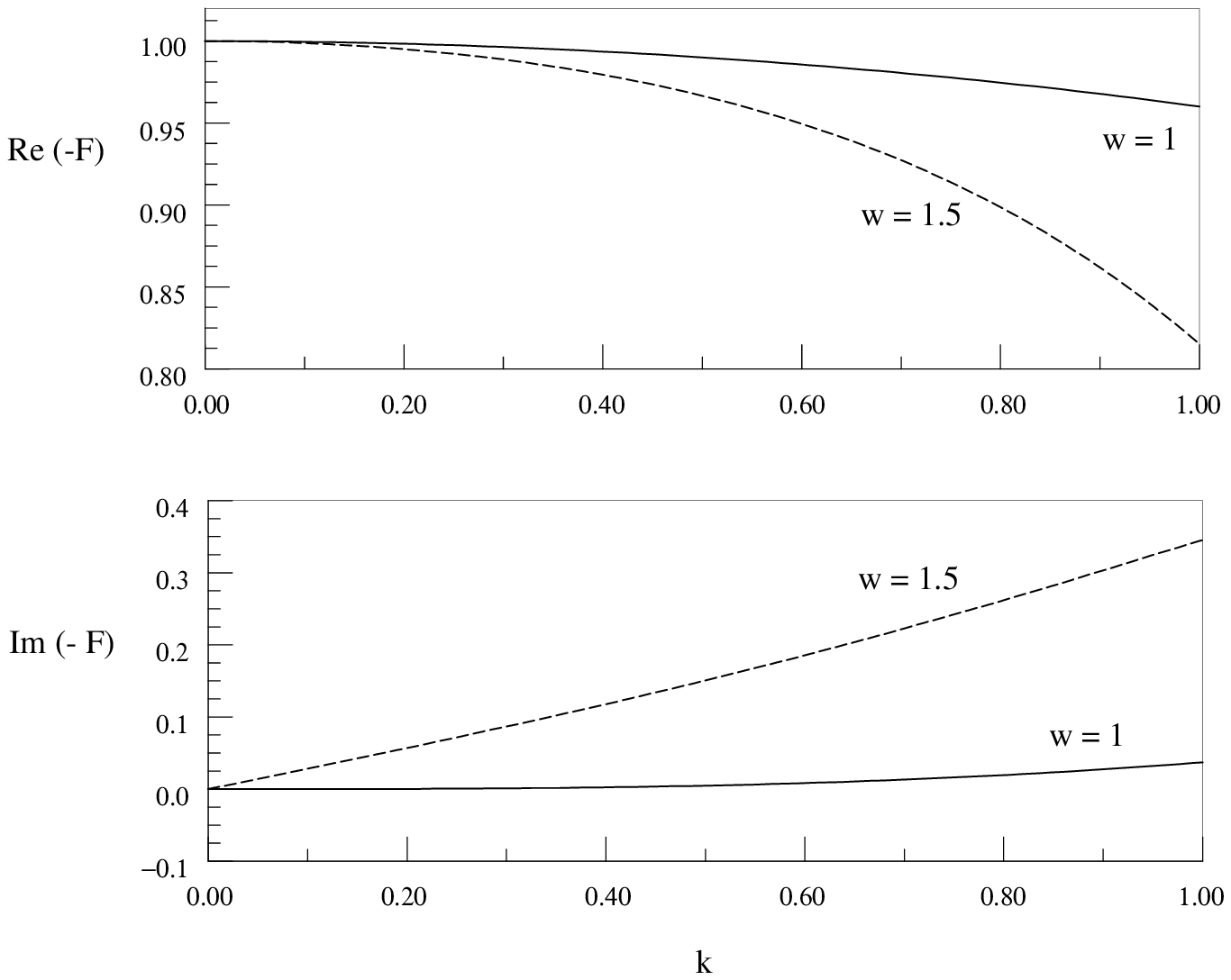}
\vsize=5 cm
\caption{\sl Real and imaginary parts of the coefficient {\sl -F}
of eq.(\ref{psi}) as a function of {\sl k} for the case
$\frac{k'_0~w}{2.25~\pi}=0.9998902939413$ w=1, and 1.5 times this value, w=1.5}
\label{fig3}
\end{figure}

In figures 2 and 3 we have shown the real and imaginary parts
of {\sl F} for different well depths. Figure 2
corresponds to the case close to a transmission resonance at $k=0$, 
$\frac{k'_0~w}{2~pi}=~1.00658424209$ and twice this
value, with$k'_0=\sqrt{2~m~|V_0|}$. 
Figure 3 corresponds
to in between two transmission resonances at threshold  $\frac{k'_0~w}
{2.25~\pi}= 0.9998902939413$ and one and a half times this value.
Note however that the terminology of transmission resonance refers
to a nonzero value of {\sl k}. There is no transmission
at zero momentum.
In the actual packet scattering process, even if $k'_0~w$ is equal
to the value of a transmission resonance or in between resonances, $k'$ differs
even by a small amount from the threshold value. Therefore our pictures
display the expected behavior of {\sl F}.
Even if for a specific value of k the condition of a transmission resonance
is met, this set is of measure zero as compared to the continuum of
values of {\sl k'} entering the calculation of the reflected wave.
 
We can see from the figures that in either case, close
to a transmission resonance or in between resonances, the
drop in  $-F$ scales as $\frac{1}{w}$. The same is true for
any other choice of the product $k'_0~w$.
 Therefore we can estimate $\ds k_{max}\approx(2~w)^{-1}$. 
With a proportionality factor that depends on the depth of the potential
only. As it is our aim to understand the absence
of a polychotomous wave train for a wide packet and not
the exact cut-off, we proceed
with the above estimate as a working hypothesis.
For the imaginary part of the argument in the $sin$ function in eq.(\ref{asymp})
 to stay small and not blurr up completely the pattern we need
$\ds 2~k_{max}~\s^2~q_0<<1$, or, using our scaling argument

\bea\label{constraint}
\s<<\sqrt{\frac{w}{q_0}}
\eea

For a fixed incoming momentum, the 
initial packet width has to be smaller than a certain proportion
of the well width. This was indeed observed in numerical simulations\ci{pra}.
The interference between incoming and reflected waves {\sl D} and {\sl F}
terms in eq.(\ref{psi}) is responsible for the diffractive pattern.
The transmitted wave has no such two components. There is no possible
interference and no diffraction peaks may be observed in
the transmitted packet.

All the above arguments pertain only to the specific example
presently considered. However, the example carries
the essential ingredients for any type of packet. One only needs
to replace the form factor $e^{-z}$ in eq.(\ref{asymp}), by the corresponding
expression for any other chosen packet.
It does not depend on the type of potential. Other functional dependencies
may give more complicated expressions than the simple {\sl sin}
function above, but essentially they will show the same structure.

The all important contribution to the diffraction
in space and time comes from the interference between
the incoming packet and the 
reflected wave originating from momenta $k\ra 0$. The latter amounts
to the excitation of a virtual state at almost 
zero energy quasi-bound state inside the well,
 as was indeed observed numerically in \ci{pra,jpa}.
The extreme importance of zero energy metastable states inside
the well, is recognized in the
literature for quite some time. In the limit of $k=0$, 
the virtual state becomes a half bound state.
This state is termed usually a half-bound state
due to a factor of one half in the phase shift at
threshold in the derivation of Levinson's theorem in
the nonrelativistic case\ci{sassoli}, 
and the relativistic case of the Dirac equation.\ci{dong}.
Its existence affects the phase shift at threshold. 
The present work shows that the existence of metastable
almost zero energy virtual states
makes itself evident through wave packet diffraction in space and time.
The effect should be independent of the sign of the potential.
Transforming to a barrier is like transforming a
slit to an obstacle (Babinet's principle). Both show diffraction
in space. Therefore, we should see diffraction in space and time
for barriers too, provided the width of the barrier is smaller
than the initial width of the packet.
This is indeed the case as shown in figures 4 and 5.
\begin{figure}
\epsffile{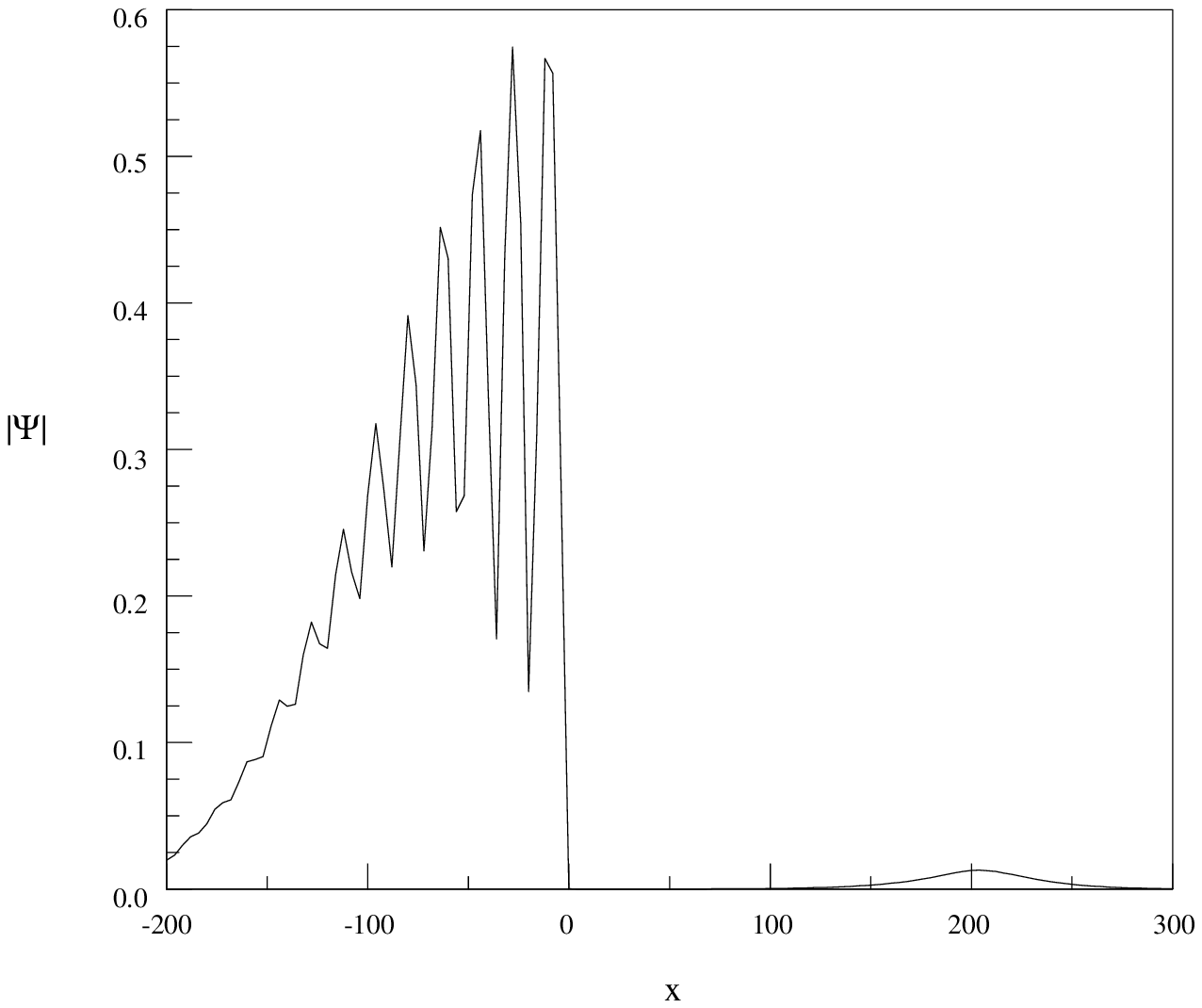}
\vsize=5 cm
\caption{\sl $|\Psi|$ as a function distance x for an initial
wave packet of width $\s=0.5$ starting at $x_0$=-10 impinging upon a barrier
of width parameter w=1 and height $V_0$=0.2 at t=800, the initial
average momentum of the packet is q=1.}
\label{fig4}
\end{figure}
\begin{figure}
\epsffile{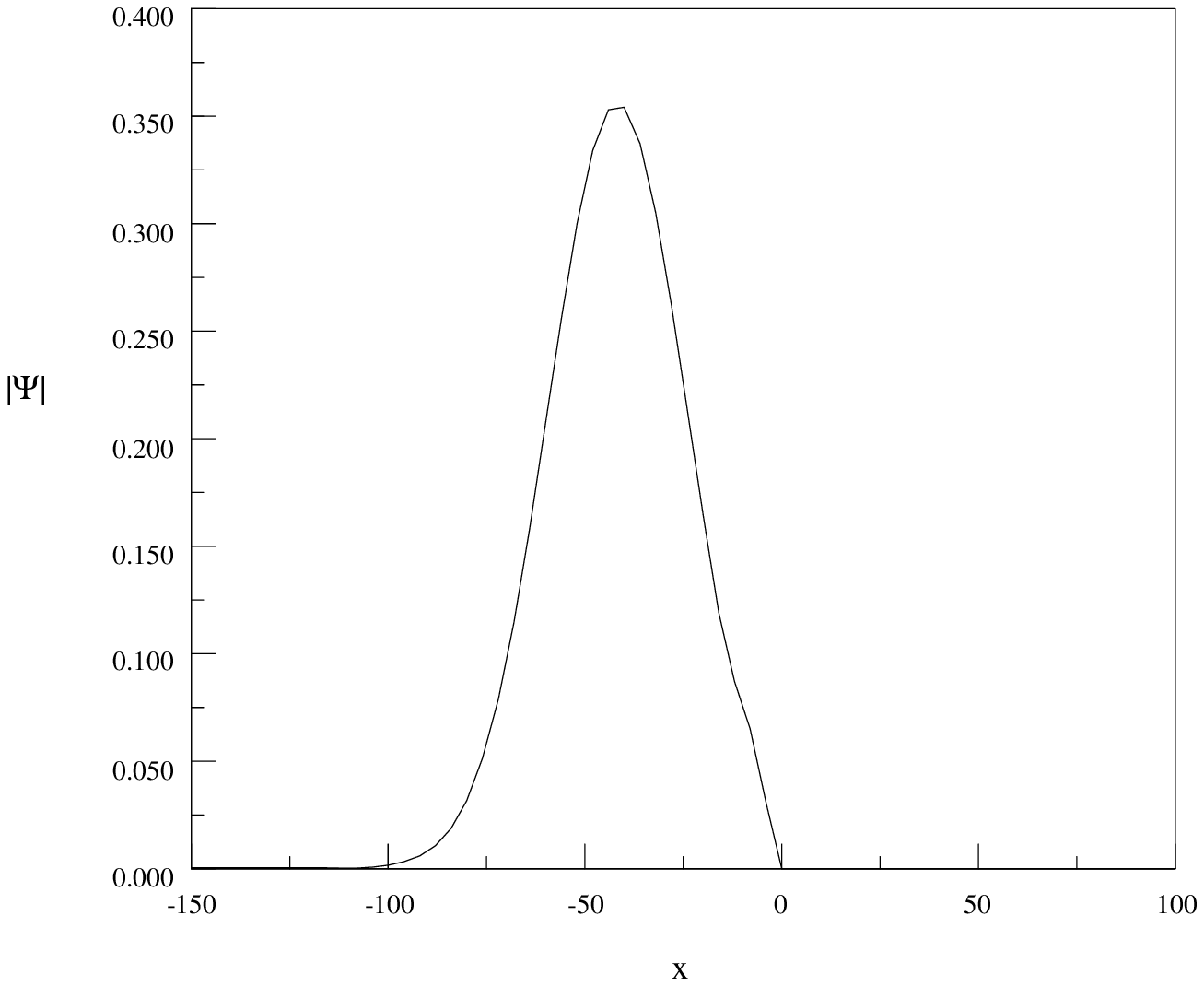}
\vsize=5 cm
\caption{\sl $|\Psi|$ as a function distance x for an initial
wave packet of width $\s=2.$ starting at $x_0$=-10 impinging upon a barrier
of width parameter w=1 and height $V_0$=0.2 at t=1200, the initial
average momentum of the packet is q=1.}
\label{fig5}
\end{figure}

A minimal uncertainty gaussian wave packet traveling from the left
with an average speed $v$, initial location $x_0$, mass {\sl m}, wave number
$q~=~m~v$ and width $\s$,

\bea\label{packet}
\psi~=~C~exp\bigg({i~q~(x-x_0)-\frac{(x-x_0)^2}{4~\s^2}}\bigg)
\eea
is scattered from a Gaussian barrier

\bea\label{barrier}
V(x)=V_0~exp\bigg({-\frac{x^2}{w^2}}\bigg)
\eea

Let us now compare to the phenomenon of {\sl diffraction in time}\ci{mosh},
 and {\sl diffraction in space and time} with plane waves\ci{
zeilinger, felber}.
The former arises when a shutter is
suddenly opened at t=0. Behind the shutter there is 
a stream of monoenergetic particles.
After the opening of the shutter, there arise
oscillations in the current on the other side of it.
These oscillations die out at large times.
The latter is produced by the combined effect of
shutters that open in time and interfere between them in space, or
other variations\ci{zeilinger}. 
Instead of having a vanishingly small time correlation length\ci{gahler},
the process of wave packet diffraction in space and time, 
persists for infinite time.
However, for the polychotomous pattern to survive we need
an initial spread of
the packet smaller than the extent of the potential.
Translating to a framework of time scales, this
would imply that the time scale of the opening of the
shutter should be longer than the time scale generated from the width of the
packet, namely $\ds\sl \Delta t_{shutter}
>\frac{\s~m}{q_0}$, where, as above,
$\s$ is the initial width of the packet.
It appears interesting to find out whether
a finite opening time, may prevent the exponential decay
of the diffraction in time process also.

\section{\sl Diffraction in space and time for the Dirac equation}

The diffraction in time phenomenon\ci{mosh}, does not
exist in the relativistic case. This can be seen from eqs. (18) and (34)
in ref.\ci{mosh} for the ordinary wave equation and the Klein-Gordon
equation respectively. Only when the speed of light is taken to infinity, 
the nonrelativistic results are recovered, as can be seen
from eq.(36) in the same work.
This seems quite surprising because, the Klein-Gordon (KG) equation 
at low momenta compared to the mass, yields
very similar phenomena as the corresponding Schr\"odinger (Sch)
equation, as evidenced by a 
substitution $\ds \psi_{KG}=e^{-i~m~t}~\psi_{Sch}$.
It is not possible to open a shutter in zero time, as
it implies an infinite opening speed.
A shutter that opens slow enough 
 may provide a framework in which not only the pattern persists
to long times, but allows its existence
for the relativistic case in a causal manner too.

In order to see that wave packet diffraction in space and time does
exist in the relativistic case, we performed numerical calculations
 using the one-dimensional Dirac equation with a scalar potential
\footnote{ The use
of a potential in the non-stationary case for relativistic equations is
questionable. However, it is still of relevance.} 
 {\sl S(x)} as given in eq.(\ref{barrier})

\bea\label{Dirac}
\bigg[i\gamma^{\mu}~{{\dd_{\mu}}}+\big(m+S(x)\big)\bigg]\Psi=0
\eea

For the initial packet we took a minimal uncertainty relativistically
invariant wave packet

$\Psi_0 =\left( \begin{array}{c}U \\ V\end{array} \right)$

\bea\label{dirpacket0}
U&=&\int{dk~e^{i~k(x-x_0)-2~\s^2~(E~E_0-k~q_0-m^2)}}\nono
V&=&\int{dk~\frac{i~k}{E+~m}~e^{i~k(x-x_0)-2~\s^2~(E~E_0-k~q_0-m^2)}}
\eea
with $E=\sqrt{k^2+m^2},~E_0=\sqrt{q_0^2+m^2}$, and we have taken 
$\bf \ds \gamma_0=\bf \sigma_z~,\gamma_1=\sigma_z~\sigma_x$ for the Dirac 
matrices, where $\sigma$ denotes the corresponding Pauli matrix.

We extended the numerical method used for the nonrelativistic case\ci{gold},
 to the unitary evolution of the Dirac particle and found it extremely
accurate. This method is a straightforward extension
of the method of \ci{gold}.\footnote{A numerical algorithm and 
routines are available from the author}.
The vector density $\ds \rho=\int~dx~(|U|^2+|V|^2)$ is conserved
to an accuracy of more than 0.1\% after millions of iterations. 
The equation is
solved with high accuracy as determined not only by substitution, but,
also by comparing to the free evolution of the wave packet above at large times.
The price one has to pay for the Dirac case is a much smaller time
increment because of the oscillations introduced by the mass parameter.
Figures 6 and 7 show the results obtained.

\begin{figure}
\epsffile{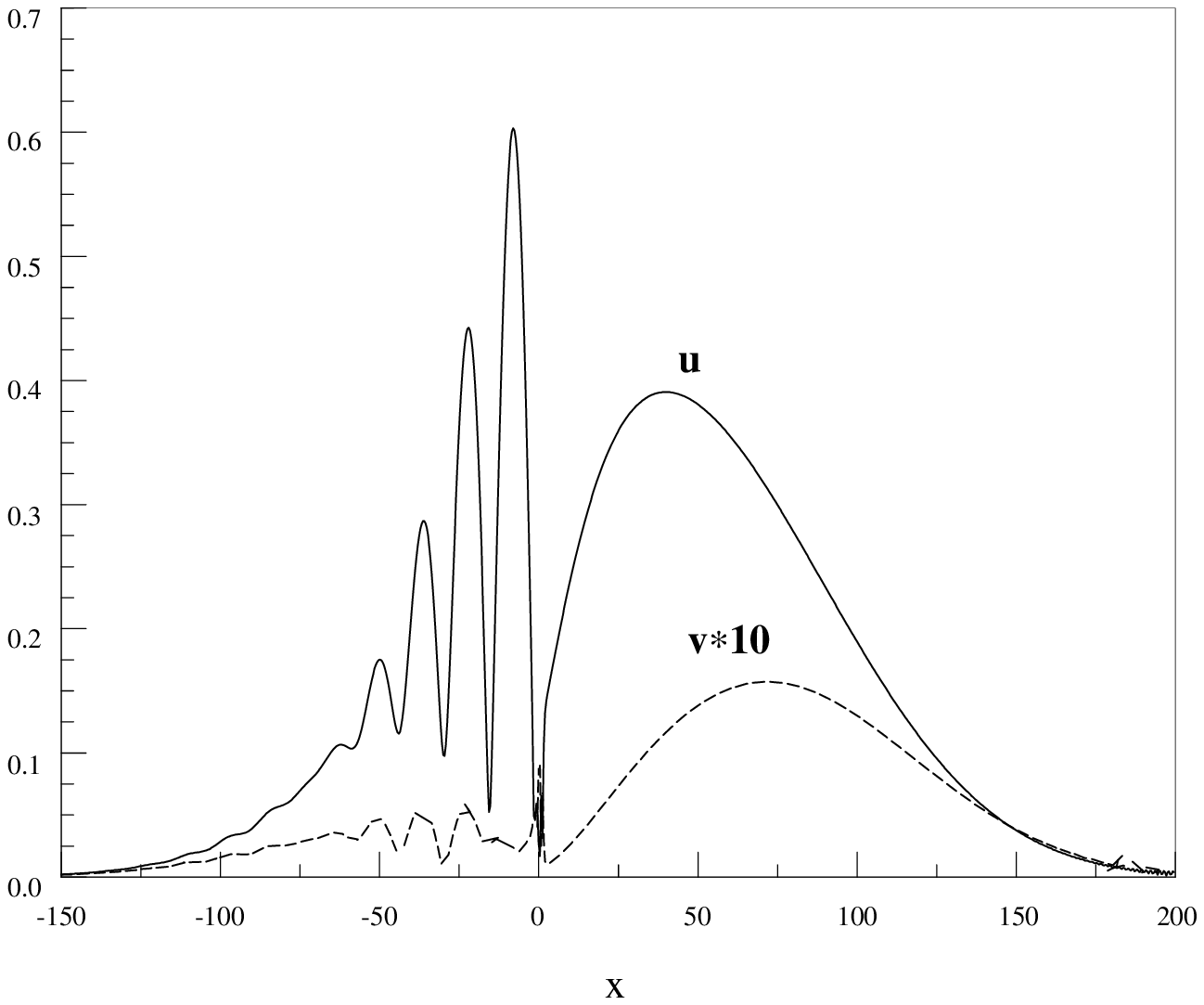}
\vsize=5 cm
\caption{\sl Upper {\bf u}=$|U|$ and ten times lower component {\bf v}=$|V|$,
of the Dirac wave function for an initial
wave packet of width $\s=0.5$ starting at $x_0$=-10, $q_0$=1
on a well of width parameter w=1 and depth $V_0$=1 at t=800}
\label{fig6}
\end{figure}
\begin{figure}
\epsffile{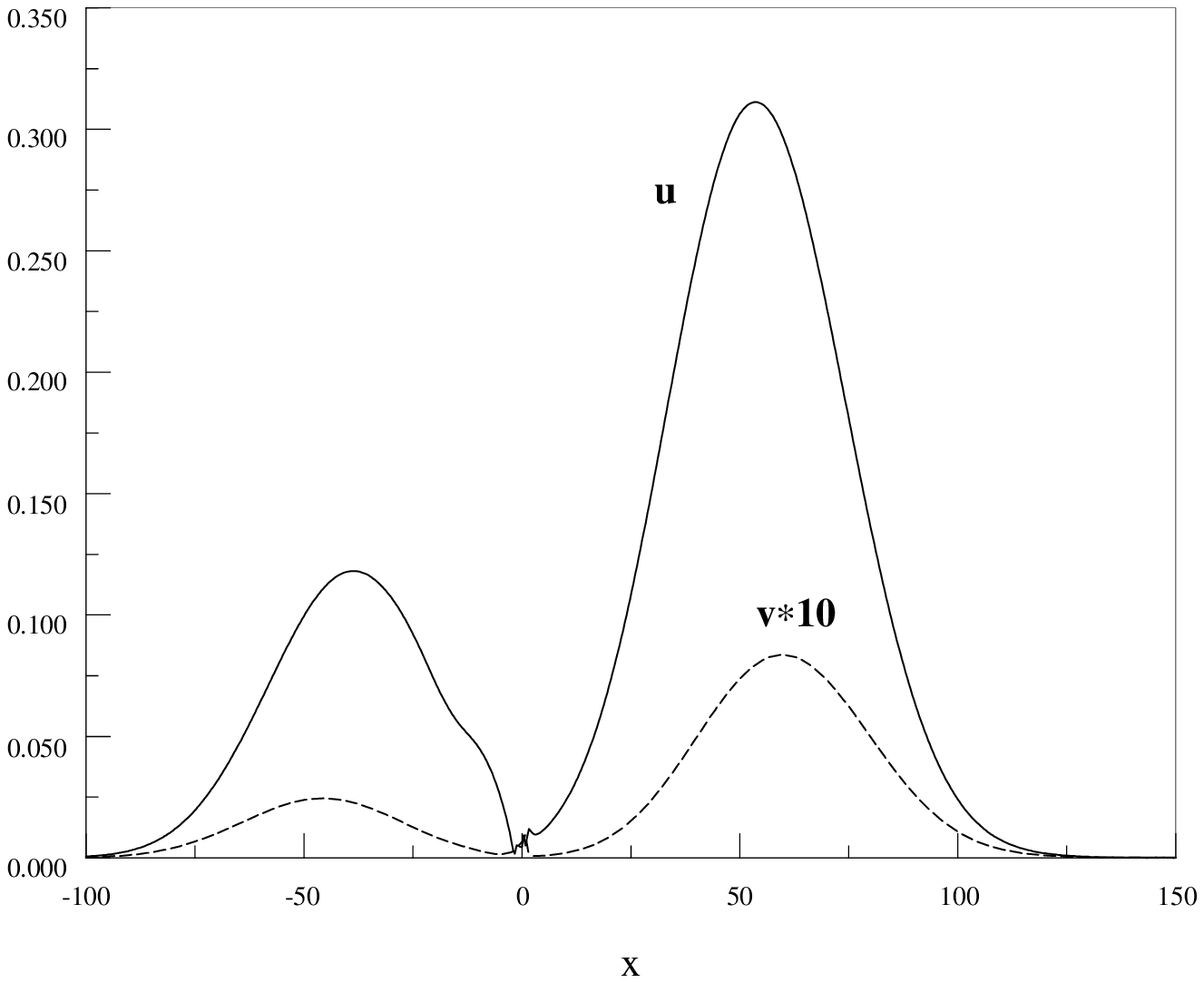}
\vsize=5 cm
\caption{\sl Upper {\bf u}=$|U|$ and ten times lower component {\bf v}=$|V|$,
of the Dirac wave function for an initial
wave packet of width $\s=2$ starting at $x_0$=-10, $q_0$=1
on a well of width parameter w=1 and depth $V_0$=1 at t=1200}
\label{fig7}
\end{figure}
It is clear from the figures that the phenomenon exists in the Dirac case
also.

We now proceed to study the analytical behavior of the 
diffraction process for the Dirac equation.
Consider the one-dimensional Dirac equation with a scalar potential only
 {\sl S(x)} as given by eq.(\ref{Dirac})
For the initial packet we use a minimal uncertainty relativistically
invariant wave packet of eq.(\ref{dirpacket0})
The stationary wave function inside and outside the well for fixed momentum
is given by $\phi(k,x,t)
 =\left( \begin{array}{c}\phi_1~e^{-i~E~t} \\ \phi_2~e^{-i~E~t}
\end{array} \right)$

\bea\label{psidir}
\phi_1(x<-w,k,t)&=&~(e^{i~k~x}+~B~e^{-i~k~x})\nono
\phi_2(x<-w,k,t)&=&\frac{i~k}{E+m}~(e^{i~k~x}-~B~e^{-i~k~x})\nono
\phi_1(-w<x<w,k,t)&=&~(C~e^{i~k'~x}+~D~e^{-i~k'~x})\nono
\phi_2(-w<x<w,k,t)&=&~\frac{i~k'}{E+m^*}~(C~e^{i~k'~x}-~D~e^{-i~k'~x})\nono
\phi_1(w<x,k,t)&=&~F~e^{i~k~x}\nono
\phi_2(w<x,k,t)&=&~\frac{i~k}{E+m}~F~e^{i~k~x}
\eea
with

\bea\label{dirpar}
B&=&\frac{(1-g^2)(e_3^2-e_4^2)}{\Delta}\nono
C&=&\frac{-2~e_1~e_4~(1+g)}{\Delta}\nono
D&=&\frac{2~e_1~e_3~(1-g)}{\Delta}\nono
F&=&\frac{-4~g}{\Delta}
\eea

where $\ds e_1=e^{i~k~w},~e_2=e^{-i~k~w},~e_3=e^{i~k'~w},~e_4=e^{-i~k'~w}$,
 $\ds~g=\frac{k'~(E+m)}{k~(E+m^*)}$, $\ds \Delta=e_3^2~(1-g)^2-e_4^2~
(1+g)^2$, $\ds k'=\sqrt{E^2-(m^*)^2},~m^*=m-V_0$.\footnote
{For a vector potential the $^*$ is in the energy instead of the mass}

The solution to the scattering of the packet becomes
$\Psi(x,t) =\left( \begin{array}{c}U(x,t) \\ V(x,t)\end{array} \right)$

\bea\label{dirpacket}
U(x,t)&=&\int{dk~\phi_1(k,x,t)
~e^{-i~k~x_0-2~\s^2~(E~E_0-k~q_0-m^2)-i~E~t}}\nono
V(x,t)&=&\int{dk~\phi_2(k,x,t)
~e^{-i~k~x_0-2~\s^2~(E~E_0-k~q_0-m^2)-i~E~t}}
\eea

As in section 2 we consider the long time behavior of the 
wave. Here too, the reflected wave will receive contributions mainly
from low values of k.
Inspection of the reflection parameter {\sl B} in eq.(\ref{dirpar}), we
find that for $k\ra~0$, $B\ra~-1$.
The same happened in the nonrelativistic treatment of section 2.
For low initial momenta, $q_0<<m, or~v<<1$, the exponential of 
eq.(\ref{dirpacket}) becomes
$\ds \approx~e^{-~ik~x_0-i~m~t-i~\frac{k^2}{2~m}t-\s^2~(q_0-k)^2}$.
So except for the factor $\ds e^{i~m~t}$, that is independent of {\sl k},
we are back at the nonrelativistic case for the upper component, and at the
same time the lower component vanishes.
Hence, the Dirac case in one dimension is exactly analogous to the 
nonrelativistic case for low average momenta of the initial wave packet.

In the ultrarelativistic limit $\ds m\ra~0$, the upper and lower components are
identical up to a factor of {\sl i}. 
The incoming and reflected waves do not spread. Hence, there
is no interference between both. 
If we first find the wave in the backward direction as in section 2, and then
take the limit of a small mass, we find that the pattern disappears, as
the arguments of the {\sl sin} function tend to zero.
For $t\ra\infty$, the backwards scattered wave
 becomes a receding wave packet located
around $x=-x_0-t-w$, without any diffraction. The incoming packet is no
longer present in the region behind the well.

In summary, the diffraction in space and time
with wave packets exists in the relativistic regime, but gradually
diminishes as the mass decreases, until 
is eventually washed out completely for the massless case, as was
 found for the {\sl diffraction in time} process with the
Klein-Gordon equation.\ci{mosh}

\section{\sl Wave packet diffraction in three dimensions}

In section 2 it was shown analytically that
there exist a broad class of phenomena occurring in nonrelativistic and
relativistic quantum wave packet potential scattering
that behave as time dependent {\sl persistent} diffraction patterns.
The patterns are produced by a time independent well or barrier.
Differing from the {\sl diffraction in time}\ci{mosh} and
{\sl diffraction in space and time}\ci{zeilinger} phenomena, 
that apply to the time dependent opening of slits with
 plane monochromatic waves, the pattern does not decay exponentially with time.

We now consider the generalization of the analytical formula
 of eq.(\ref{asymp}) to three dimensions.
We take a gaussian wave packet in three dimensions,

\bea\label{packt}
\phi(\rr,t)~&=&~e^{y_0}\nono
y_0&=&{i~\qq_0\cdot~(\rr-\rr_0)-\frac{(\rr-\rr_0)^2}{4~\s^2}}
\eea

with $\qq_0$, the initial average momentum of the packet, $\rr_0$,
 the initial location and $\s$ the packet width, impinging on a spherically
symmetric square well

\bea\label{sqwll}
V(x)= -V_0~\Theta(w-r)
\eea

where, {\sl w} is the well  width, and $V_0$, the depth, and $\Theta$, the
step function.
The results are not dependent on the choice of wave packet
however, the Gaussian profile facilitates the integrations.
The total wave function with outgoing boundary condition at large
distances, reads\ci{gol}

\bea\label{psio}
\psi^+(\rr,t)~&=&\psi_{in}(\rr,t)+\psi_{scatt}(\rr,t)\nono
&=&\int{~d^3k~f(\kk,\qq_0)~(\psi_{in}(\kk,\rr,t)+\psi_{scatt}(\kk,\rr,t))}
\eea

where

\bea\label{psip}
\psi_{in}(\kk,\rr,t)&=&e^{i~\kk\cdot~\rr}\nono
\psi_{scatt}(\kk,\rr,t)&=&\frac{e^{i~k~r}}{k~r}\sum_L{(2~L+1)~e^{i\delta_L}
~sin(\delta_L)~P_L(\cx\cdot\ck)}\nono
f(\kk,\qq_0)&=&e^{-(\kk-\qq_0)^2~{~\s^2}-i~\kk~\cdot
\rr_0-i~\frac{k^2}{2~m}~t}
\eea

The incoming wave may be found explicitly to be

\bea\label{psin}
\psi_{in}&=&\bigg(\frac{\pi}{it/(2m)+\s^2}\bigg)^{\frac{3}{2}}~e^{y_1}~
erfc(-i\frac{D}{\sqrt{\s^2+it/(2m)}})\nono
y_1&=&-\frac{D^2}{4~ (i~t/(2m)+\s^2)}-q_0^2~\s^2
\eea

Where $D=|\rr-\rr_0-2~i~\s^2\qq_0|$, {\sl erfc} represents the complementary
error function of complex argument and the absolute value in the
definition of {\sl D} pertains to the vectorial character, and not to the
complex number. Error functions and similar Fresnel integrals are
found in many diffraction formulae, such as those appearing in the
works of Moshinsky.\ci{mosh}

Performing the angular integral over $\ck$, the scattered wave becomes

\bea\label{scatt}
\psi_{scatt}(\rr,t)&=&\int{k^2~dk~\Phi(\rr,k,t)}\nono
\Phi(\rr,k,t)&=&\sqrt{4\pi}~\frac{e^{i~k~r}}{k~r}
~e^{-\s^2(k^2+q_0^2)-i~t\frac{k^2}{2m}}\sum_{L,M}{(2~L+1)~e^{i\delta_L}
sin(\delta_L)I_{L,M}(k,\qq_0,\rr_0},\rr)
\eea

Where 
\bea\label{il}
I_{L,M}(k,\qq_0,\rr_0,\rr)
&=&(4\pi)^2\sum_{l,l',m,m'}i^{l'-l}~C_{l,L}~j_l(k~x_0)~j_{l'}(-2i\s^2q_0k)
Y_L^M(\cx)~{Y_l^m}^*(\cx_0)~{Y_{l'}^{m'}}^*(\cq_0)\nono
C_{l,L}&=&\frac{\sqrt{(2~l+1)~(2~l'+1)}}{\sqrt{(2~L+1)}}
<l~l'~L|0~0~0~>~<l~l'~L|m~m'~M>
\eea

Here $Y_l^m$ represents as usual the spherical harmonics, $j_l$ denotes
the spherical Bessel function for complex argument, and $<l~l'~L|...>$ 
represents the Clebsh-Gordan coefficient.

The expression for $I_L$ is quite involved. However, for $L=0$ it
simplifies considerably.
Fortunately, this is the only contribution we need at long times,
 except for very specific cases for which the phase shift of higher partial 
waves dominates.
Fot $t\ra\infty$ the integral in eq.(\ref{scatt}) oscillates wildly, except for
values of {\sl k} that give an extremal phase\ci{bohm}.
It is easy to show that in this case, only extremely small values of k
enter the integral. However, for very small values of {\sl k}, 
reads\ci{gol}

\bea\label{phase}
tan({\delta}_l)&=&-\frac{(k~w)^{2~l+1}}{(2~l-1)!!~(2~l+1)!!
}~\frac{z_l-l}{z_l+l+1}\nono
z_l&=&\frac{x~j'_l(x)}{j_l(x)}
\eea

where $\ds x=k'~w=\sqrt{k^2+2~m~|V_0|}~w$, 
and $j_l$, denotes the spherical Bessel function of order {\sl l}.
Except for very specific cases for which there is a resonance 
at a higher partial wave, only
the lowest partial wave matters.
Higher partial waves are suppressed by factors of the form 
$\ds\bigg(\frac{w}{\sqrt{it/(2m)+\s^2}}\bigg)^l$. 
For very long times we can then approximate the scattered wave by taking only
the $L=0$ term in eq.(\ref{scatt}).
For this case $I_L$ becomes

\bea\label{I0}
I_0&=&4\pi~j_0(k~d)\nono
d&=&|\rr_0+2~i~\s^2\qq_0|
\eea

where again the absolute value pertains to the vectorial character
of the variable inside the vertical bars, and not to the complex number.
Therefore, $j_0$ is really a complicated mixture of Bessel and
Hankel functions. 

Inserting eqs.(\ref{I0},\ref{phase}) in eq.(\ref{scatt}) and recalling the
rules for the integration of the error function for complex variables, 
we find the transparent result
\bea\label{scatt1}
\psi_{scatt}&=&-\frac{u_0}{d}
\bigg(\frac{\pi}{it/(2m)+\s^2}\bigg)^{\frac{3}{2}}~e^{-q_0^2\s^2}
\bigg(\frac{r+d}{2~r}~e^{\lambda_1^2}~erfc(-\lambda_1)-\frac{r-d}{2~r}~
e^{\lambda_2^2}~erfc(-\lambda_2)\bigg)\nono
\lambda_1&=&i\frac{(r+d)}{2\sqrt{~(it/(2m)+\s^2)}}\nono
\lambda_2&=&i\frac{(r-d)}{2\sqrt{~(it/(2m)+\s^2)}}
\eea

With $ u_0=w~(1-\frac{tan(\kappa~w)}{\kappa~w})$, $\kappa=\sqrt{2~m~|V_0|}$,
and {\sl d} is defined in eq.(\ref{I0}).

The incoming and scattered wave interfere in the total wave.
Noting that the scattered wave depends on the absolute value of $\rr$,
while the incoming wave depends on $\rr$ itself, it is clear
that there will be a different behavior at forward and backward angles.

For long times and long distances we can safely approximate
the complementary error functions with the value {\sl erfc}$\approx2$.
In order relate to the results of the one dimensional case, we
will bring here the expression for the full wave, in the case
of a packet impinging from a
distance $r_0$ at zero impact parameter, intial momentum $q_0$, 
on a well located
at the origin. 
Using equations (\ref{psin},\ref{scatt1}) we find

\bea\label{back}
\psi&=&2~i\bigg(\frac{\pi}{it/(2m)+\s^2}\bigg)^{\frac{3}{2}}~e^{\alpha}
~sin(\frac{m~r}{t}~(r_0+2~i~q_0\s^2)+i\alpha)~e^x\nono
e^{\alpha}&=&\sqrt{\frac{u_0~(2~d+~u_0)}{d^2}}\nono
x&=&{-q_0^2\s^2-\frac{r^2~m^2}{t^2}\s^2+i\frac{m}{2~t}~r^2}
\eea

This is a diffraction pattern that travels in time as found for the
one dimensional case. Again the pattern gets blurred and 
forms a single peak when the imaginary part of the argument of the
{\sl sin} function is large. As explained in section 2, this
amounts to the condition $\s>>\sqrt{\frac{w}{q_0}}$.
There is here the additional blurring effect
of the phase $\alpha$.
For narrow enough packets the pattern persists to infinity.

The results are independent of the impact parameter
provided the initial packet is located at a much longer distance along the
direction of $\qq_0$ as compared to the transverse direction, 
as found numerically in ref.\ci{jpa}.
Also the pattern does not depend strongly on the initial
energy, as long as it stays small compared to the well depth or barrier
height.
The formula of eq.(\ref{back}) may be used to find the angular
distribution, the scattering and total cross sections, etc.
Such expressions do not seem to exist in the literature for packets, even
for the simplest square well case dealt with here.
\begin{figure}
\epsffile{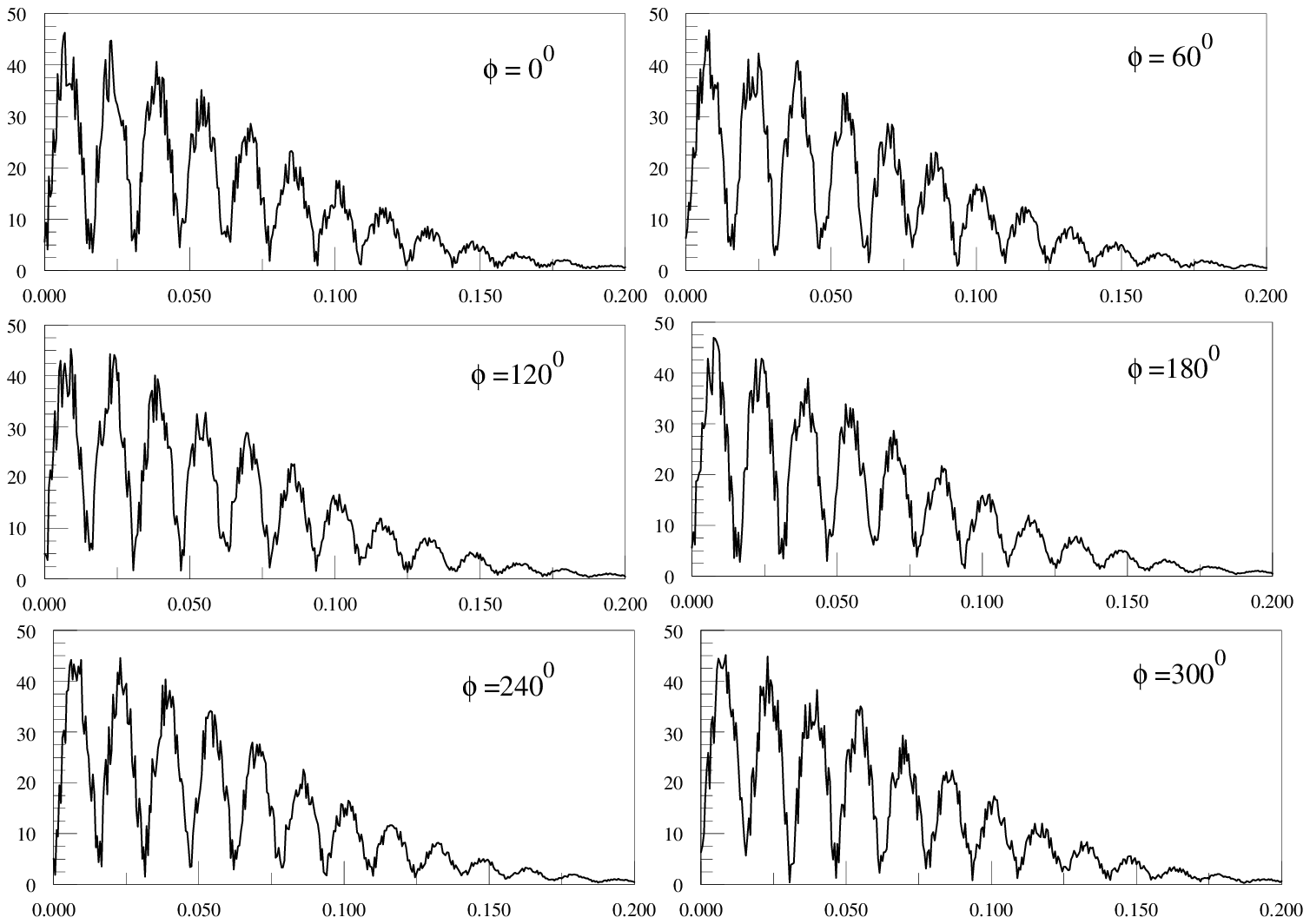}
\vsize=5 cm
\caption{\sl $|\psi|$ as a function of distance for various
polar angles $\phi$ and in the plane z=0 for three dimensional scattering. 
(See text)}
\label{fig8}
\end{figure}

The present treatment took advantage of the
properties of Gaussian packets and the simplicity of the square well.
However, they are general. One has only to replace the Gaussian form factor
by the appropriate one in case and the phase shift for
the square well by the one corresponding to the specific choice of potential.
The case of long range potentials deserves however some care.

In order to visualize somehow the three dimensional results we present
in figure 8 a hypothetical scattering event of a cold neutron packet
of 1 Fermi width, 
with initial velocity of {\sl v=0.02 c} impinging along the x-axis
on a well located around the origin of width 10 Fermi and depth 40 MeV.
The initial position of the neutrons is taken to be $x_0$ = -20 Fermi,
$y_0$ (the impact parameter)= 2 Fermi and the time is taken to be
$t=5~10^{14}$ in units of Fermi also. 
The numbers are taken just for
the sake of exemplification and are unrealistic experimentally regarding
the initial width of the packet.
For a more viable experimental setup see section 5. We chose the potential 
range and width to give a large scattered wave contribution.
Figure 8 depicts the wave amplitude 
obtained by adding equations (\ref{psin},\ref{scatt1}) as a function of
distance in meters and we have multiplied
the wave function by
a factor of $({it/(2m)+\s^2})^{\frac{3}{2}}$, in order to have
ordinates with values around $|\psi|$=1.
We have chosen to show graphs for the plane {\sl z=0}.
The pictures are similar for other planes.
The darkened elements in the figure are due to fast oscillations
of the wave.
\begin{figure}
\epsffile{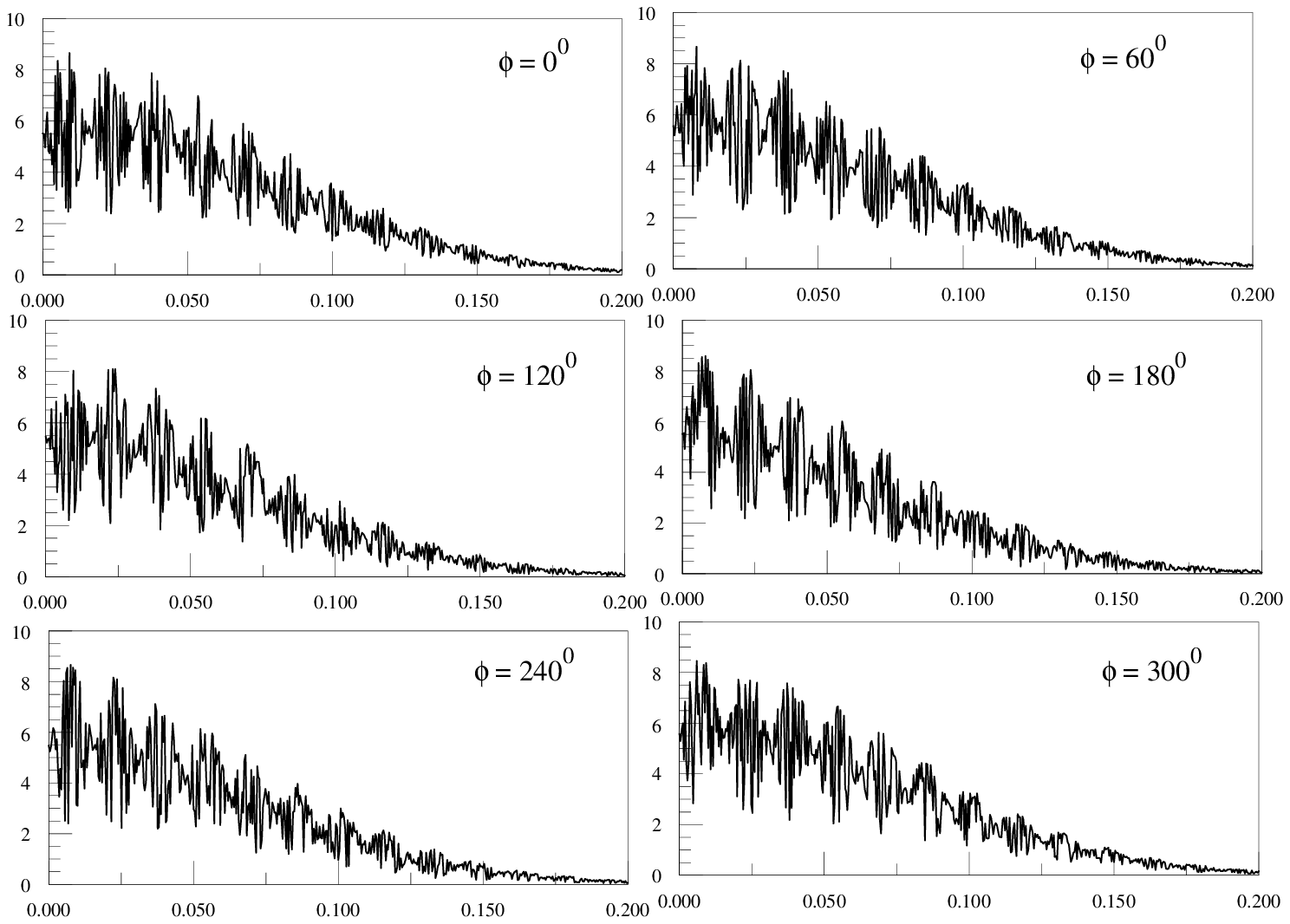}
\vsize=5 cm
\caption{\sl $|\psi|$ as a function of distance for various
polar angles $\phi$ and in the plane z=0 for three dimensional scattering. 
Well depth increased by 5 \% as compared to the previous figure}
\label{fig9}
\end{figure}
For this very advantadgeous choice of parameters the diffraction in
space and time in three dimensions exists at all angles and not
only in the backward direction. 
Perhaps this is expected for a very low speed packet that essentially
spreads around a position close to 
the origin, while a secondary scattered wave interferes
with it. Directionality is here weak, in contradistinction to the
one dimensional case.
The effect depends strongly on the input parametes.
In figure 9 we modified the parameters of figure 8 only slightly
by increasing the well depth by 5\%.
The effects are dramatic. Wild oscillations in small regions of space
arise. This topic will be studied in a later work.

\section{\sl Numerical simulation of a suggested experiment}

Present advances in bose traps permit the handling of atoms at very low
temperature and velocities. At such energies the atoms move as if they
were a wave packet provided the random agitation due to thermal effects
is not as crucial.
However, still the large amount of interactions
between the particles and of them with the environment 
causes a decoherence of the wave in quite a short time.
Instead of resorting to a cold bose gas, we will then focus
on a drop of liquid Helium.
Although the experimental details are beyond the expertise of the
author, a hypothetical setup will be described
in the hope that it is not too far away from reality.

We will imagine a setup that may be dealt with
approximately as a one-dimensional system, namely a drop of
liquid Helium put on a surface a short distance apart from an
impervious wall and far from any other boundary, inside a
adiabatic container. Gravity is supposed to hold the drop in place
vertically. The wall serves as a potential barrier and we are only
interested in the horizontal spread of the drop due to the
interaction with the wall.

Consider a drop of liquid helium, of around $1~cm^3$ in volume
lying near a corner of Cesium coated plates perpendicular to
each other may be appropriate for this purpose. 
Cesium is needed in order to prevent wetting\ci{nacher}.
The helium, plate
and environment should be  at a temperature
below the $\lambda$ point, although it would be
interesting to see the effect of normal to superfluid component
of the Helium on the wave packet behavior, as a tool to study decoherence.
Put a large area detector far behind the drop and the plate.
The number of particles measured as a function of time at a
fixed position would then be determined by the absolute value of
the wave function at that point. The particles are repelled
by the wall and the packet spreads at the same time.

Figure 10 depicts such a case. We solved the one dimensional Schr\"odinger
equation for a packet as described in section 2
, with a barrier of very large strength.
We assumed a total number of particles of $N=5*10^{21}$, obtained from
the tabulated density of liquid helium and a volume of 1 $cm^3$.
We used a plate (potential) of width w=1 cm and strength $V_0=4~eV\approx
10^{15}$, in units of $\hbar=1$. We took a detector size of $dx=1~mm$, and
recorded the amplitude of the wave multiplied by {\sl dx} in order
to calculate the relative number of particles in the same distance
and at varying time. The number of particles as a function of time
-relative number multiplied by the total initial number in the drop
for a properly normalized wave initially- is shown in the figure.

It appears that, due to the large number of particles to be detected a
simple weighing technique might be feasible, in this way we also
avoid interaction of the packet with the detector.
\begin{figure}
\epsffile{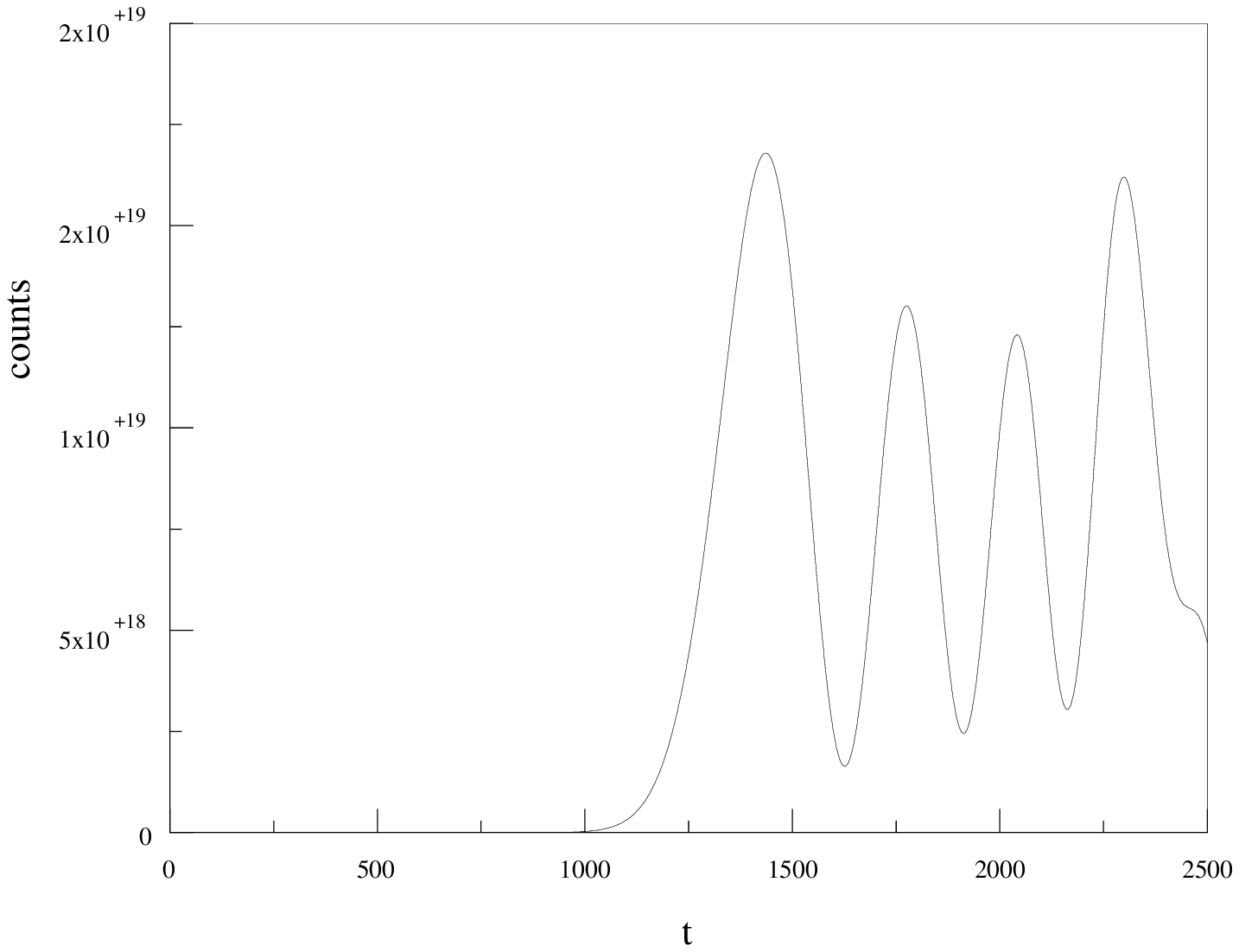}
\vsize=5 cm
\caption{\sl Number of Helium atoms counted at a distance of 5 cm
behind a plate of thickness w=1 cm as a function of time in seconds}
\label{fig10}
\end{figure}

It is perhaps too optimistic to expect the proposed experiment will work
so cleanly as in the simulation, due to all kinds of effects inside the
drop that lead to decoherence, such as production of
internal excitations like rotons, vortices, etc.
However, some remainder of the effect might still
show up in the counter.
If it does, it will be a triumph for the quantum mechanical description
of matter waves by means of wave packets of macroscopic size.

\section{\sl Summary}
In summary, we have found in the present work that, 
the phenomenon of diffraction of wave packets in space and time,
is determined by the spreading of the
incoming waves and the 
subsequent interference between incoming and scattered waves as well
as by the physical size of scatterers.
It appears then, that the theoretical and experimental study of wave
packet scattering as it evolves in time, gives us a new tool 
for the investigation of structure and dynamics of
atoms and nuclei and a new possible testground for the predictions
of quantum mechanics.

{\bf Acknowledgments}

The anonymous referee very constructive remarks, are greatly appreciated.

\end{document}